\documentclass[12pt,preprint]{aastex}
\usepackage{spr-astr-addons}

\newcommand{\HI}{H\,{\sc i}}
\newcommand{\HII}{H\,{\sc ii}~}

\newcommand{\Msunpyr}{\ensuremath{{\rm M}_{\odot}{\rm yr}^{-1}}}
\newcommand{\mum}{\ensuremath{\mu\mbox{m}}}

\begin{document}

\title{Modeling IR Spectral Energy Distributions:  \\ A Pilot Study of Starburst Parameters and Silicate Absorption Curves for Some GOALS Galaxies}
\shorttitle{Modelling IR SEDs of Spitzer GOALS Galaxies}
\shortauthors{Dopita et al.}

\author{ Michael A. Dopita\altaffilmark{1}\altaffilmark{2}, Lee Armus\altaffilmark{3}, Lisa J. Kewley\altaffilmark{2}, Jeff A. Rich\altaffilmark{2},  Dave Sanders\altaffilmark{2}, Phillip N.  Appleton\altaffilmark{4}, Ben H. P. Chan\altaffilmark{5}, Vassilis Charmandaris\altaffilmark{6}, Aaron S. Evans\altaffilmark{7}, David T. Frayer\altaffilmark{4}, Justin H. Howell\altaffilmark{3}, Hanae Inami\altaffilmark{3}, Joseph A. Mazzarella\altaffilmark{5}, Andreea Petric\altaffilmark{3}, Sabrina Stierwalt\altaffilmark{3} \& Jason Surace\altaffilmark{3} }
\email{jrich@ifa.hawaii.edu}
\altaffiltext{1}{Research School of Astronomy and Astrophysics, Australian National University, Cotter Rd., Weston ACT 2611, Australia }
\altaffiltext{2}{Institute for Astronomy, University of Hawaii, 2680 Woodlawn Drive, Honolulu, HI 96822, USA}
\altaffiltext{3}{Spitzer Science Center, MS 220-6, California Institute of Technology, Pasadena, CA 91125, USA}
\altaffiltext{4}{NASA Herschel Science Center, California Institute of Technology, MS 100-22, Pasadena, CA 91125, USA}
\altaffiltext{5}{Infrared Processing and Analysis Center, California Institute of Technology, MS 100-22, Pasadena, CA 91125, USA}
\altaffiltext{6}{University of Crete, Department of Physics, Heraklion 71003, Greece ; IESL/Foundation for Research and Technology-Hellas, GR-71110, Heraklion, Greece}
\altaffiltext{7}{Department of Astronomy, University of Virginia, P.O. Box 400325, Charlottesville, VA 22904, USA}

\email{Michael.Dopita@anu.edu.au}

\begin{abstract}
This paper describes a pilot study into the spectral energy distribution (SED) fitting and the derivation of physical parameters for 19 galaxies observed as part of the Great Observatories All-sky LIRG Survey (GOALS) survey as observed with the \emph{Spitzer Space Telescope}. For this we have used the pan-spectral fitting tools developed in a series of papers by Dopita and his co-workers. We show that the standard Lee and Draine `astronomical silicate' model cannot provide a good fit to the silicate absorption features as observed in the heavily dust-extinguished ($A_{\rm V} \sim 50$mag.) starbursts. We have derived an empirical fit to the `starburst silicate' absorption in these objects. This absorption curve is consistent with the silicate grains being systematically larger in starburst environments than in the local Galactic interstellar medium. We demonstrate the sensitivity of the SED fitting to each of the fitted parameters, and derive these parameters for those galaxies which do not have an embedded AGN. This technique is simple and provides reasonably robust and uniform parameters for the starburst, especially as far as the star formation rate, population of old stars, compactness of the starburst region and total foreground extinction are concerned. However, the chemical abundances and the optical extinction cannot be reliably determined by this analysis, and optical SEDs will also be required to provide a complete characterization of the starburst region and of the surrounding galaxy.

\end{abstract}

\keywords{ISM: dust --silicates---extinction, \HII---galaxies:general,star formation rates, spectral energy distributions, starburst-infrared:galaxies}

\section{Introduction}
The spectral energy distribution (SED) of a galaxy is the observational quantity needed to derive most of the the fundamental parameters of that galaxy. From its study we can hope to determine the star formation rate, the quantity of older stars, and their star formation history, the amount of dust and its nature, and the chemical composition of both the stellar and interstellar components \citep{Dopita05a,Walcher10}.

For starburst galaxies, a wide variety of empirical, semi-emirical and purely theoretical techniques have been applied to the study of their SEDs, For a recent comprehensive review see \citet{Walcher10}. In the semi-empirical modelling of Dale and his collaborators \citep{Dale01, Dale02}, the SEDs of both disk and starburst galaxies were taken as a one-parameter family in terms of the dust temperature, with more active star formation being related to higher mean dust temperature.  \cite{Lagache03} derived an empirical  sequence of SEDs in terms of the absolute luminosity of the galaxy. 

Both of these semi-empirical approaches are valid to some extent, since IR luminous galaxies generally have greater rates of star formation, more compact star formation regions and higher luminosities than normal galaxies. Both of these models have found their greatest application in providing a simple one-parameter fit to the SEDs of normal disk galaxies.

In starburst galaxies - to which we now confine our discussion - the geometry of the star forming region is very important in determining the form of the IR SED at the wavelengths observed by the \emph{Spitzer Space Telescope}. The SED in this wavelength region is dominated by the emission from the hotter dust grains located in and around the \HII regions associated with the young stellar clusters, and by the polycyclic aromatic hydrocarbon (PAH) emission located in the dense photodissociation regions (PDRs) surrounding these \HII regions.  PAHs are destroyed by ionizing photons, so cannot exist in the \HII regions, but they require energetic UV photons absorbed in the 2200\AA\  band (and shortwards) to excite the characteristic C-H, C-C and C-C-C skeletal deformation modes of the PAH IR bands \citet{Draine03}. This spatial segregation of the PAH emission into the PDR is very clearly seen in the \emph{Spitzer Space Telescope} image of M20  by \citet{Rho06}. The spatial scale of the PDR regions from which most of the IR emission originates is fairly small. For example, around an \HII region with $P/k \sim 3\times 10^6$cm$^{-3}$K, typical of starbursts \citep{Kewley01}, a PDR with $A_V \sim 3$ and $T \sim 1000$K \citep{Tielens85} is only $ \sim 0.2$pc thick!

The star formation regions of starburst galaxies are both denser and of higher pressure than the more extended gas and dust associated with the older stars. The truth of this can be simply demonstrated by considering the \citet{Kennicutt98} star formation law, which has been shown to apply both to normal disk galaxies as well as starburst galaxies. This states that the local surface rate of star formation, $\Sigma_{\rm SFR}$, is related to the surface density of gas, $\Sigma_{\rm gas}$, by a simple power law; $\Sigma_{\rm SFR} \propto \Sigma_{\rm gas}^{1.5}$. Thus, by definition, high local specific star formation rates will always be associated with dense, high density regions with both high dust temperatures and high foreground extinction. However, it does not necessarily follow that the IR luminosity and the dust temperature are simply correlated, since the luminosity scales as the total star formation rate, while the the average dust temperature is determined by the average geometry of individual star formation regions. For example, a galaxy with many individual \HII regions spread through a disk may have a high total star formation but such a galaxy would have  a lower dust temperature than one in which the same star formation rate is crushed into a dense nuclear starburst.

The simplest geometries simply co-mix the star formation and the dust \citep{Devriendt99}. A somewhat more sophisticated approach was taken by \citet{Cunha08}, who used the \citet{Charlot00} approach to dust absorption and emission. This recognises that the younger stars are more dust-embedded than the older stars, and that the `birth cloud' emission is hotter than the more `diffuse ISM' dust emission. However, the dust emission templates are rather simple, consisting of a PAH template and variable gray-body contributions.

Spherical radiative transfer models have been much employed in the literature. This approach was pioneered by Rowan-Robinson and his co-workers \citep{RR80,RR89,RR93} and developed to a high degree of sophistication by  \citet{Takagi03a,Takagi03b} and by Seibenmorgen and his team(\emph{see} \citet{Siebenmorgen07} and references therein). These have been applied to the determination of star formation rates in Starbursts.

The difficulty with all of these models is that they do not contain significant \HII region physics. This leads to inconsistencies in the pressure and density of the gas, since the equilibrium temperature is not explicitly solved for. Furthermore, they cannot reproduce the hot dust seen only in \HII region. Thirdly, they fail to account for nebular line and continuum emission processes. In starbursts, the \HII regions around clusters expand dynamically during the stellar lifetime. \citet{Efstathiou00} made models which explicitly followed the evolution of the giant molecular clouds owing to the ionization-induced expansion of the \HII regions and the evolution of the stellar population, following the Bruzual \& Charlot  stellar population synthesis models (\emph{see} \citet{BC03} for more up-to-date models). In these models, the effect of transiently heated dust grains/PAHs on the radiative transfer, as well as multiple scattering, is taken into account. However, there is no gas physics.

Dopita, Groves and colleagues have developed a versatile technique to model the pan-spectral energy distributions of starburst galaxies. These methods cover a wide range of metallicities and physical parameters described briefly below. These models and their physical assumptions have been fully described in a series of papers \citep{SED1,SED2,SED3,SED4}, and will be only briefly recapitulated here. The reader is encouraged to read \citet{SED4} in particular, since the models given there are used in throughout the remainder of this paper.

These models take into account all the atomic gas physics and radiative transfer through the \HII regions and their surrounding PDRs in the outward-only approximation. The effect of multiple scattering is only included through approximate scattering-length methods within the \HII regions. The gas ionization state and temperature, and the grain temperature distribution and emission are explicitly computed at each radial step. All relevant emission lines and nebular continuum processes are all fully taken into account. The model dust has  silicates, carbonaceous grains as well as PAHs. The dust composition reflects the chemical composition of the gas. The amount of dust is calculated using solar depletion factors, and the size distribution of the grains is a power-law appropriate to a grain-shattering distribution. The optical properties of the grains are taken from \citet{DL84,WD01,Li01,Draine03a} and \citet{Draine03b}. The grain size spectrum is distributed into  80 bins for each grain type, and the emission spectrum of transiently heated dust grains is computed following the methods of \citet{GD89} and \citet{DL01}. For the PAHs, the opacity is set according to the \cite{Li01} approximation, and the ionization state and heating of the PAHs is explicitly computed. The absorbed UV/visible radiation is re-radiated into a multi-component standard PAH emission band template in the mid-IR.

Each \HII\ region +PDR model provides the expected SED of an individual \HII\ region at a chosen pressure, excited by a particular cluster, with a particular age. The art of our global SED modeling is to sum these models together so as to provide  the self-consistent evolution of a whole ensemble of \HII regions of different ages and cluster masses within a dense high pressure region of the interstellar medium (ISM). The dust temperature distribution which, for a particular choice of the metallicity, primarily determines the form of the IR SED, was shown to be set by a \emph{Compactness Parameter}. The concept of such a parameter was first developed by  \citet{Takagi03a} and \citet{Takagi03b} in the context of simple spherical radiative transfer models. In our case  the compactness parameter $\cal C$ is defined in terms of the mean stellar cluster mass, ${M_{\rm cl}}$, and the pressure in the ISM, ${P_{0}}$, by:

\begin{equation}\label{eqn:Cparam} 
\log{\cal C}=
\frac{3}{5}\log\left[ \frac{M_{\rm cl}}{M_{\odot}}\right] +
\frac{2}{5}\log\left[ \frac{P_{0}/k}{{\rm cm}^3{\rm K}} \right]
\end{equation}

The complete set of parameters which enter into the SED fitting are described in the following section. In this  paper we will apply these modeling techniques spectra drawn from a sub-set of the The \emph{Great Observatories All-Sky LIRG Survey} (GOALS) galaxies to investigate the silicate absorption properties in starburst regions, and to self-consistently derive star formation rates and other physical parameters for those galaxies which do not contain an embedded AGN.

\section{Modelling the Starburst SED}
The SED fits developed in \citet{SED1,SED2,SED3} and \citet{SED4}are based upon the stellar SEDs and the computed stellar mechanical energy input from the STARBURST 99 code \citep{Leitherer99}. We use a 1-D hydrodynamic code and our MAPPINGS III code to calculate the emission line, PAH, dust and nebular continuum contributions. We also use the \citet{Fischera03,Fischera04,Fischera05} turbulent screen models to compute the foreground attenuation at all wavelengths.

The \citet{SED4} models used in this paper are based on the 2006 version of the STARBURST 99 code. We have used a \citet{Kroupa02} broken power-law IMF between 0.1 and 120 $M_{\odot}$. Within the code, we use the standard combination of the Geneva and Padova stellar evolution tracks \citep{Vazquez05} to tabulate the mechanical energy delivered by the stars, both as stellar winds and as supernova explosions. In our SED fitting, the mechanical energy input is  used to solve the 1-D hydrodynamical evolution of the \HII regions, given a mean cluster mass and pressure in the ISM. Our MAPPINGS III code is then used to solve for the dust continuum, nebular continuum and emission line spectrum at each age, and the resulting spectra are then summed over all ages. To this \HII region spectrum we then add the spectra of ultra-compact \HII regions associated with single stars, a dense circum-starburst molecular gas screen, the spectrum of older  ($10 \leq t \leq 100$~Myr) stars which we assume to surround the starburst region, and finally a foreground turbulent attenuating dusty screen  \citet{Fischera03,Fischera04,Fischera05}. These models are starburst-only in the sense that we do not include a stellar population more ancient than 100 Myr.

The objective of this modeling was to provide a geometry which is reasonably realistic, which correctly accounts for the evolution of the ensemble  of \HII regions, and which is characterized by as few physical variables as possible. Indeed, our final SED models are characterized by only seven parameters:
\begin{itemize}
\item{The current star formation rate, SFR. This is the total star formation rate in stellar clusters averaged over the last 10Myr. This is a free parameter which determines the absolute scaling of the SED models.}
\item{The metallicity of the ISM: $Z/Z_{\odot}$. This is computed for the abundances of the STARBURST 99 models: $0.04, 0.1, 0.2, 0.4, 1.0$ and $2.0 Z/Z_{\odot}$\citep{Leitherer99}.}
\item{ The compactness parameter $\cal C$, defined above. This is computed in steps of 0.5 over the range $4 \leq \log{\cal C} \leq 8$.}
\item{The fraction, $F_{\rm UCHII}$, of young ($< 1.0$Myr) ultra-compact \HII\ (UCHII) regions and their PDRs surrounding individual massive stars \citep{Dopita06c}.  A value $f_{\rm UCHII}=1.0$ implies that within the first 1.0Myr of evolution, the fraction of star formation rate producing UCHII regions is the same as the formation rate of stars in the massive clusters, \emph{i.e} that 50\% of \emph{all} massive stars younger than 1.0~Myr are isolated and surrounded by ultra-compact \HII\ regions.}
\item{The optical extinction, $A_{\rm V(SB)}$, associated with a dense turbulent foreground screen of dusty molecular and \HI\ gas surrounding the main star formation complex \citep{Fischera03,Fischera04,Fischera05}.}
\item{The fraction, $F_{\rm old}$ of an older starburst stellar population with ages $10 \leq t \leq 100$~My. This allows for the possibility that star formation has been ongoing for at least a dynamical timescale. A value of $f_{\rm old}=1$ corresponds to steady star formation, \emph{i.e.} that  the star formation rate in the period $10 \leq t \leq 100$~Myr is the same as in the  period of $0 \leq t \leq 10$~Myr. }
\item{The optical extinction, $A_{\rm V(old)}$, of a foreground turbulent attenuating dusty screen \citep{Fischera03,Fischera04,Fischera05}.}
\end{itemize}

\section{The GOALS Data}
GOALS combines data from the NASA \emph{Spitzer Space Telescope, Chandra X-Ray Observatory, Hubble Space Telescope (HST)}, and the \emph{Galaxy Evolution Explorer (GALEX)} observatories, together with ground-based data, into a comprehensive imaging and spectroscopic survey of over 200 low-redshift ($z < 0.088$), Luminous Infrared Galaxies (LIRGs)\footnote {\emph{see} {\textsf http://goals.ipac.caltech.edu}}. 

These LIRGs represent a complete subset of the IRAS Revised Bright Galaxy Sample (RBGS), which comprises 629 extragalactic objects with 60\mum\ flux densities above 5.24 Jy, and Galactic latitudes above five degrees, and span the full range of galaxy types, AGN, and galaxy merger phases which are able to produce objects so bright at far-IR wavelengths. The full GOALS sample is listed by \citet{Armus09}, where redshifts, luminosity distances and infrared luminosities, $L_{\rm IR}$ are given using a cosmology with $H_0 = 70$ km~s$^{-1}$ Mpc$^{-1}$, $\Omega_{\rm V}= 0.72$, and  $\Omega_{\rm M}= 0.28$. Throughout this paper we adopt the cosmological parameters H$_{0}$=70.5~km~s$^{-1}$Mpc$^{-1}$, $\Omega_{\mathrm{V}}$=0.73, and $\Omega_{\mathrm{M}}$=0.27, based on the five-year WMAP results \citep{Hinshaw09}.

In a recent paper, \citet{Howell10} has compared the UV and far-IR properties of these galaxies. These authors find extreme specific star formation rate (SSFR) in the GOALS, with a median value ($3.9\times10^{-10}$ yr$^{-1}$) - comparable to the highest SSFRs seen in the Spitzer Infrared Nearby Galaxies Survey sample. The GOALS LIRGs on average have greater IR excesses than would be expected based on their UV color, consistent with very large extinctions in the main star forming regions. Both of these properties are consistent with the GOALS sample representing the most extreme star formation environments in the local universe.

In another recent paper \citep{U10} have studied the SEDs for 65 (U)LIRGs drawn from the GOALS sample, and spanning the entire electromagnetic spectrum from X-ray to radio wavelengths. The objective of this work is different from our study. \citet{U10} used stellar population synthesis models to estimate the integrated stellar masses of these systems, and the \citet{Dale02} templates to estimate the total IR luminosity, and to study a variety of color indices. Our objective in this paper is to test the starburst SED models, and to use these to derive what physical parameters we can from the IR spectroscopy obtained with the \emph{Spitzer Space Observatory}.

The galaxies presented here are sources for which the full Spitzer Infrared Array Camera (IRAC) photometry and complete Infrared Spectrograph (IRS) spectroscopy \citep{Houck04} have been obtained. The IRS instrument was used in both its high- and low- resolution modes. The wavelength bands covered by each of the four channels are 5.3-14.2 \mum, 10.1-19.4 \mum, 14.3-36.3 \mum, and 19-36.8 \mum. The observed  wavelengths were corrected to the wavelength in the rest-frame of the galaxy observed, and the fluxes were corrected to absolute $\nu F_{\nu}$ flux units (erg~cm$^{-2}$s$^{-1}$) using the luminosity distance. 

The data used here simply represents an exploratory subset of the GOALS galaxies, and is in no sense complete or even representative of the full GOALS dataset. We have simply selected the sources to be analyzed in this paper from amongst the 50 galaxies in the GOALS First Data Delivery. In order to ensure a large fraction of starburst-dominated sources, the galaxies were also selected to have strong PAH features. ÊDouble sources were deliberately excluded to avoid possible contamination by a companion. ÊSelection of sources also avoided the obviously extended sources for the same reasons of beam-correction or confusion. In this sense, the galaxies presented here are the morphologically more ``simple" sources in the sample, and are therefore presumably the most amenable to a simple integral SED analysis. 

All the IRS spectra were extracted using the standard pipeline, and flux calibrated assuming a point source. Thus these spectra represent "nuclear" spectra in all cases, and are not intended to provide an integrated MIR spectrum of the LIRG. In fact, many of the observed LIRGs are somewhat resolved to IRS \citep{Diaz-Santos09}.  This means that since the slits are narrow, slit losses (and gains) are difficult to quantify.  In some cases therefore, the orders do not match up at the boundaries of the spectra (at the 10\% level).  For point sources this cannot happen, but for extended sources it does because the \emph{Spitzer Space Telescope} is diffraction limited and the spectrograph slits are narrow.  The absolute flux calibration is based on point sources (stars), so that we automatically take into account slit and diffraction losses in the pipeline, but only for a point source. As a result, it was sometimes necessary to normalize the spectra relative to each other. The low resolution data was helpful in this regard, since it overlaps with the higher resolution spectra. In the spectra presented here, we have taken the normalization wavelength to be 25\mum, since this corresponds to the IRAS observation wavelength, thus facilitating comparison with integrated fluxes measued with this instrument.  Furthermore, the IRAC photometry had to be extracted and summed in a region which corresponds as closely as possible to the IRS slit. Both of these steps have been taken for the data presented in this paper. In fitting the SEDs, both the high and the low resolution IRS spectroscopy is used. However, the resolution of the high resolution spectra is higher than the resolution of the models, and the observed line to continuum contrast is therefore higher than in our our model fits.

\section{Fitting the IR Spectra}
The IR fitting procedure is not a linear process, as each parameter affects a particular part of the IR spectrum, and in different ways. In what follows we will demonstrate this and show how most of the SED fitting parameters can be derived in a fairly uncoupled way. However before we discuss the detailed fitting, let us first discuss the issue that was encountered in first attempting to fit the silicate absorption features at the nominal wavelengths of $\sim 10$ and 18\mum .

\subsection{Silicate Absorption}

\begin{figure}
\includegraphics[width=0.9\hsize]{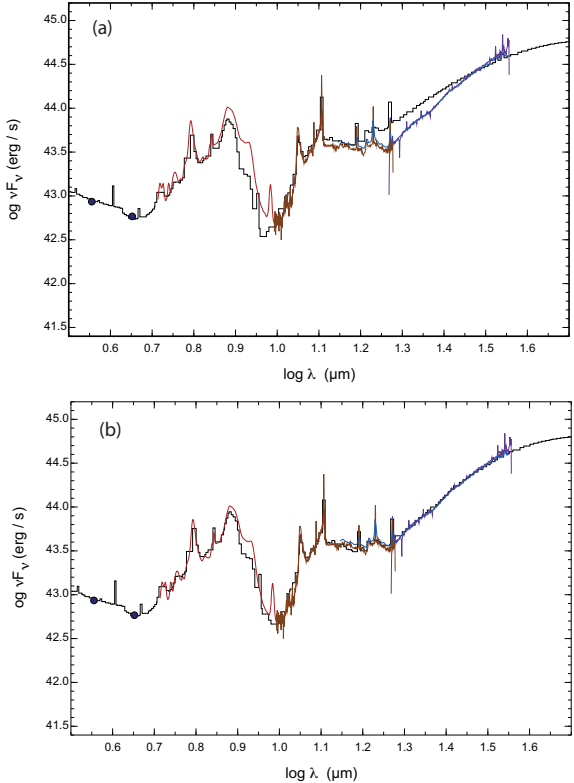}
\caption{The comparision of the SED fit to ESO 507-G070 using (a) ``astronomical silicates'' as defined by \citet{DL84} and (b) the modified ``starburst silicate'' absorption derived in this paper. The colored lines represent the IRS spectra, the blue points the IRAC photometry points, and the stepped black line (histogram) is our SED model fit at each of the wavelengths computed in our model.}\label{fig_Si_fit}
\end{figure}
\begin{figure}[here]
\includegraphics[width=0.9\hsize]{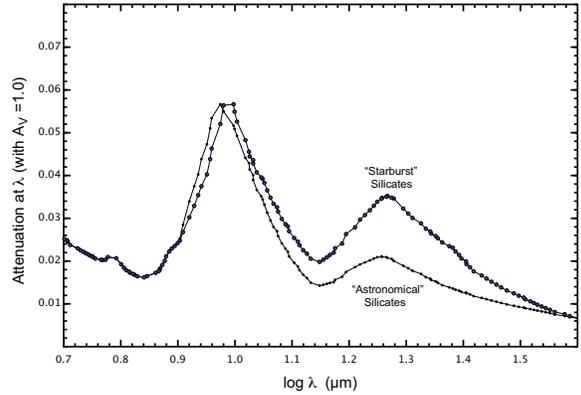}
\caption{The comparison of the total ($\log_{10}$) attenuation produced by a grain mixture including the ``astronomical silicates'' as defined by \citet{DL84} and one including the modified ``starburst'' silicate absorption derived in this paper. This was obtained by empirically fitting the observations of the ULIRG galaxy ESO 507-G070 maintaining the normalization at the peak of the10\mum\ absorption feature. Note that the ``starburst silicate'' curve shows a shift of the 10\mum\ feature to longer wavelength, a more pronounced and shifted 18\mum\ absorption feature, and a shoulder in the absorption at about 24\mum. All of these are characteristic of larger Si grains (\emph{c.f.} \citet{Dorschner95}, figure 9). The absorption feature at $\log \lambda \sim 0.78$\mum\ is a water ice feature.}\label{fig_Si_Opacity}
\end{figure}

Our models - like most others- used the \citet{Li01} ``Astronomical Silicate'' scattering and absorption properties. The optical constants used in the  \citet{Draine03a,Draine03b} model do not depend on the laboratory measurements of any particular silicate material or any particular size distribution. Rather, they are empirically derived to fit the shape of the silicate absorption seen the the Galactic neighborhood. In our fitting, we rapidly discovered that these parameters do not provide an adequate fit to either of the two absorption bands in the most heavily reddened objects with $A_{\rm V} > 50$mag. 

As an example, we show in panel (a) of Fig \ref{fig_Si_fit} a fit to ESO 507-G070 (IRAS 13001-2339). This fit implies a foreground extinction of  $A_{\rm V} \sim 8$mag. and a starburst extinction of  $A_{\rm V} \sim  52$mag. It is clear that the fit with ``Astronomical Silicate'' absorption fails in the following ways:
\begin{itemize}
\item{The peak of the 10\mum\ feature is too deep, and at the wrong wavelength.}
\item{The 18\mum\ feature is not deep enough, is of the wrong shape, and its peak is also at the wrong wavelength.}
\end{itemize}

In order to better fit the silicate features, we also took an empirical approach, and adjusted the absorption coefficients by hand until we were able to reproduce the depth, shape, and central wavelengths observed for silicate features in ESO 507-G070 (IRAS 13001-2339). In order to avoid changing the normalization, we kept the central absorption of the $\sim 10$\mum\ feature constrant. The resulting fit is shown in panel (b) of Fig \ref{fig_Si_fit}. This implies a foreground reddening of  $A_{\rm V} \sim 8$mag. and a starburst extinction of  $A_{\rm V} \sim 48$mag. All spectral features in the spectrum are now well fit by this ``Starburst Silicate'' fit.

The resulting absorption coefficients for our grain mixture are graphed in Figure \ref{fig_Si_Opacity} and are listed for general reference in Table \ref{tab:Si-abs}. Note the shift of the peaks of the two absorption features to longer wavelengths (9.57\mum\ and 18.44\mum\ respectively, instead of 9.11\mum\ and 18.08\mum\ ), the much stronger 18\mum\ absorption relative to the 10\mum\ feature, and the appearance of a ``shoulder'' on the absorption at 24\mum . 

The form of the silicate absorption depends upon a number of variables. Both  smooth structure of the observed silicate bands is an indication that the dust is highly amorphous rather than having a well-defined crystal structure. A consideration of the cosmic abundances and the condensation temperatures implies that the Mg and Fe silicates, such as amorphous olivines and pyroxenes is preferrred \citep{Ossenkopf92}. The original \citet{DL84} ``astronomical silicate'' absorption curve has a small 18\mum\ to 10\mum\ absorption band ratio of 0.37. According to \citet{Ossenkopf92}, values this low can be produced in pure Fosterite (Mg$_2$SiO$_4$) or in irradiated olivine. By contrast, high ratios of the 18\mum\ to 10\mum\ absorption can be produced by different compositions, such as a high mass fraction of Fe$_2$SiO$_4$ or bronzite (Mg$_{0.9}$Fe$_{0.1}$SiO$_3$. The absorption of bronzite (a form of pyroxene) also displays the additional shoulder at 24\mum\  \citep{Ossenkopf92}. 

Using IR spectra of some of the ULIRGs observed by \citet{Spoon07} and \citet{Imanishi07} with the \emph{Spitzer Space Telescope} and performing radiative transfer using the DUSTY code by \cite{Ivezic99},  \citet{Sirocky08}  have used the silicate features to help constrain the dust chemistry. These observations also require dust with relatively high 18\mum\ to 10\mum\ absorption ratio of the silicate features (around 0.5), and these authors favor cold dust model of \citet{Ossenkopf92} (rather than their warm dust model, or the ``astronomical'' silicate model of Draine \citep{DL84,WD01}, since thermal gradients in the dust layer can naturally produce the observed scatter in the ratio of the 18\mum\ to 10\mum\ absorption ratio and the \citet{Ossenkopf92} cold dust model gives a high base value of this ratio. However, we have some problems with this interpretation. Whilst it may produce a change in  the 18\mum\ to 10\mum\ absorption ratio between low-density ISM environments typical of disk galaxies, and the high density environments characteristic of starbursts, it cannot at the same time explain the systematic shift to longer wavelengths of both of the silicate absorption profiles which we have found here \citep{Dorschner95}.

A simple explanation is that the grains in starburst environments are systematically larger than in the local ISM. Indeed, the rapid growth of grains in dense and cool molecular clouds is inevitable. The denser the environment, the more rapid will be the rate of grain growth, and we would expect  that grains will grow to a size limit imposed by the depletion of the heavy elements from the gas phase. If the molecular clouds are cool enough, grains can also grow by coagulation (although such grains will be very weak, and easily shattered if they find themselves in a warmer environment).

The optical constants and the effect of size on olivine and pyroxene glasses have been carefully measured in the laboratory by \citet{Dorschner95}. From this work it is found that for olivine glass particles, as their size increases, the 10\mum\ feature initially shifts to longer wavelengths and then rapidly broadens when $a>1.25$\mum. At the same time, the ratio of the 18\mum\ to 10\mum\ absorption rapidly increases, the peak is shifted to longer wavelengths, and an absorption shoulder at 24\mum\ becomes more pronounced \emph{see}  \citet{Dorschner95}, figure 9). All of these properties correspond to the differences identified between the ``astronomical silicate'' absorption curve and our derived ``starburst silicate'' absorption. To provide a 18\mum\ to 10\mum\ absorption band ratio of 0.63, and an absorption peak in the principal silicate feature at 9.57\mum\   (as observed) would imply a mean grain size of $1.0 < a < 1.5$\mum\ in the starburst environment, a not unreasonable estimate. 

Thus, in conclusion,  it seems almost certain that the change in the silicate absorption is the result of larger grain sizes in the starburst environment. However, larger silicate grains are less effective in producing optical extinction. Thus, for a given depth in the silicate features, the implied $A_{\rm V}$ will be less, but we would also expect that the implied column density of gas would be correspondingly larger. In what follows, these (uncertain) correction factors have not been taken into account in deriving the parameters of the starburst.

\subsection{Fitting the SED}
Fitting the SED to seven variables might be initially thought to be a very difficult procedure. Fortunately, however, most of these variables can be simply and unambiguously separated in these IR spectra, since they affect different spectral regions, and in different ways. In the following sub-sections we illustrate the sensitivity of the model SEDs to each of these parameters in turn, by varying the parameters enough to make a significant change to the overall SED. The one parameter which cannot be determined at all from the IR SEDs is chemical abundance. In our fitting, we have varied the abundance in the range 0.4 to 2.0 solar, but this only significantly affects the scaling of the PAH emission features. In principle, the abundance might be able to be estimated from the ratio of the PAH features to the continuum at 25 \mum\. However, the uncertainties are at present too large to make this reliable. To derive accurate abundances will require optical SEDs and in particular, optical emission line fluxes, to be obtained for these galaxies.

\begin{figure*}
\includegraphics[width=\hsize]{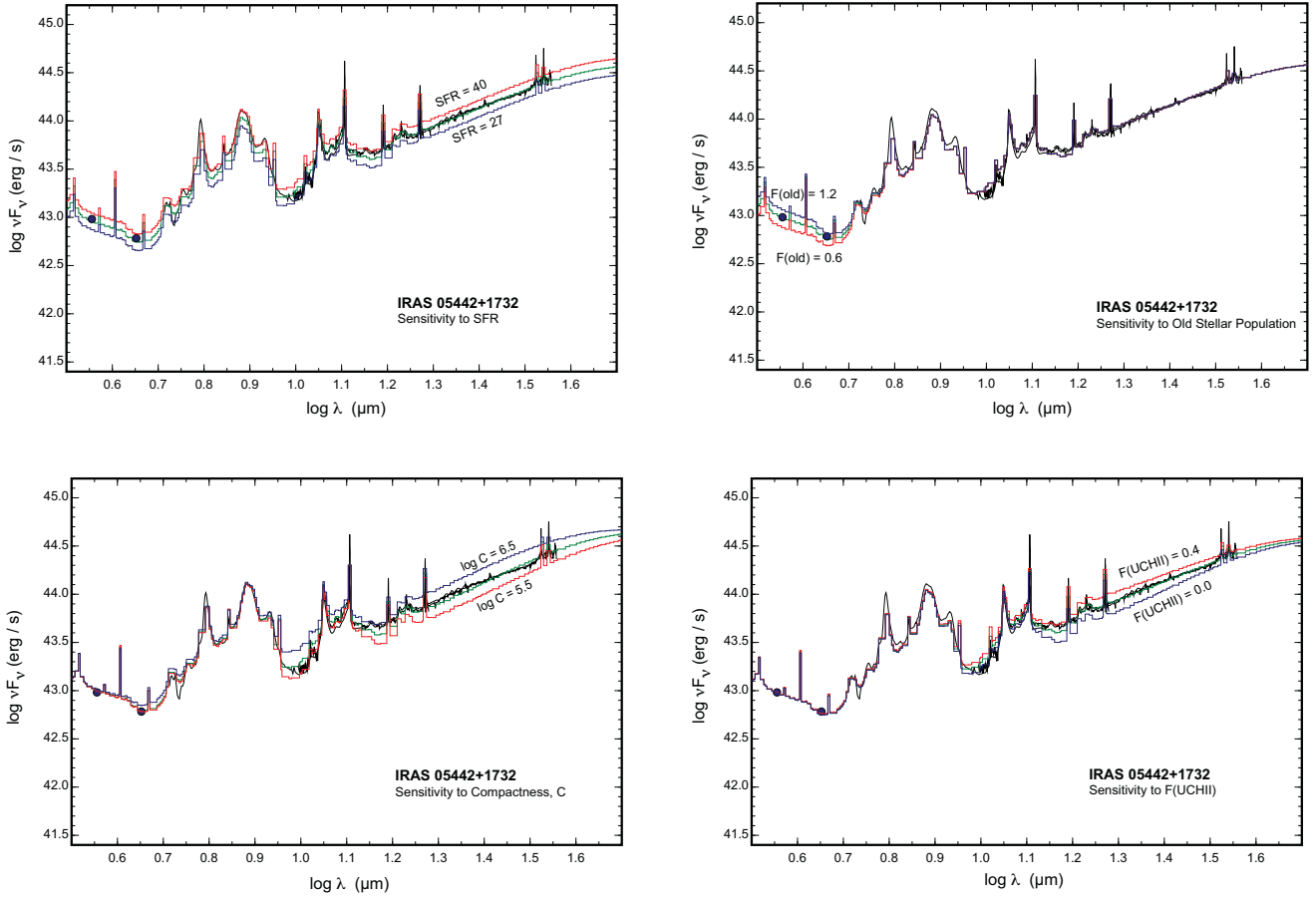}
\caption{The sensitivity of the SED fitting to particular variables is demonstrated for one galaxy, IRAS 05442+1732. The stepped colored lines represent the models, and the black lines and points are the observational data. In panel (a) the star formation rate (SFR) is shown for three values,  27, 33 and 40 \Msunpyr. In panel (b) the old stellar population fraction, $F_{\rm old}$, is given fo the values 0.6, 0.9 and 1.2. In panel (c) we vary the compactness parameter $ \log{\cal C} $ from 5.5 through 6.0 to 6.5. Finally, in panel (d) we adjust the fraction of ultra-compact HII regions,  $F_{\rm UCHII}$, in the steps 0.0, 0.2 and 0.4. }\label{fig_Sens_fit}
\end{figure*}

\subsection{Star Formation Rate}
The fundamental and most important parameter to determine is the star formation rate (SFR). In the IR, the whole starburst region acts rather as a bolometer of the luminosity of young stars, since both the PAH and the dust continua scale as the star formation rate. Changing the SFR simply scales the spectra up or down. This is shown in panel (a) of figure (3). This demonstrates that the star formation rate can be measured from these spectra (at virtually any wavelength) to an absolute accuracy of $\pm 10$\%. 

We elect to determine the star formation rate by least-squares minimization of the difference between our model and the observations at all  wavelengths observed by the IRS. The IRAC data points are not used, since these are also sensitive to the older stellar fraction, as discussed in the next sub-section. 

\subsection{Old Stars}
The fraction of older stars, $F_{\rm old}$, as mentioned above, is determined by the IRAC photometry. This works because the old stellar component only becomes significant in determining the SED at shorter wavelengths than the main PAH emission features. Panel (b) of Figure (3) shows that the fraction $F_{\rm old}$ is set by the normalization to the IRAC 3.5\mum\ and 4.5\mum\ photometry, after the underlying star formation component has been set as described above. We always find an excess flux at the IRAC wavelengths that can be ascribed to the older stellar component, and we can fix this to an accuracy of about $\pm 15$\%. For an unattenuated SED, the IRAC 3.6\mum\ to 4.5\mum\ color is fixed essentially by the slope of the Rayliegh-Jeans tail of the Black-Body distribution. As we show below, a flattening of the  IRAC 3.6\mum to 4.5\mum\ color may be used to help constrain the foreground attenuation, $A_{\rm V(old)}$, but this color index is not particularly sensitive as an attenuation measure.
\begin{figure}
\includegraphics[width=\hsize]{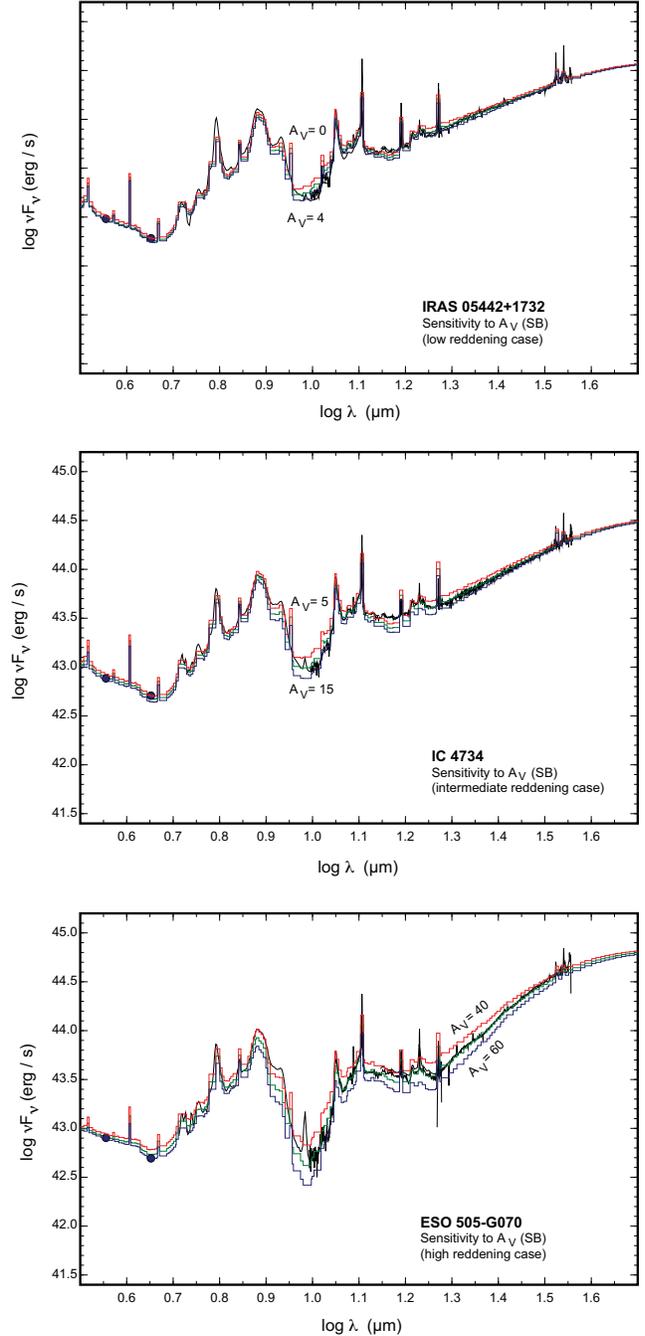}
\caption{The sensitivity of the SED in fitting to the attenuation of the starburst, $A_{\rm V(SB)}$ in the wavelength range covered by the IRAC and IRS observations. This is given for three diffferent cases, a low-extinction object, panel(a), an intermediate extinction case, panel (b) and a high-extinction case, panel (c). It is clear from the goodness of the fit in these examples that  for $A_{\rm V(SB)} < 3$mag. the attenuation is essentially unmeasurable,  while for $A_{\rm V(SB)} >5$mag. the attenuation can be estimated to an accuracy of $\sim \pm 5$mag.}\label{fig_Sens_Av}
\end{figure}

\subsection{Compactness Parameter}
The compactness parameter, $ \log{\cal C} $, a measure of the geometrical relation between the dust and the heating stars determines the temperature distribution of the dust. Higher compactness parameters move the peak of the far-IR emission to shorter wavelengths. However, at wavelengths below about 30 \mum\ the slope of the dust continuum emission is essentially unchanged as the compactness parameter is increased. However, for compactness parameters above $ \log{\cal C} > 6$ a flattening of slope in the dust continuum starts to become evident. The main effect is to increase the 20-30\mum\ continuum with respect to the PAH features, and to fill in the region between the principal PAH features at about 10 \mum. This is shown in Figure (3) panel (c).  We can probably determine $ \log{\cal C} $  to within $\pm0.3$ from a combination of the ratio of the 7\mum\ PAH features to the 9\mum\ continuum, the degree of veiling of the 10-14\mum\ PAH complex, and the slope of the 18-34\mum\ dust continuum.

\subsection{Ultra-compact \HII Regions}
The Ultra-compact \HII regions produce an SED which is strongly peaked near $\sim 20$\mum\ as a consequence of the very hot dust in the \HII region itself, which strongly competes with hydrogen for the absorption of the ionizing photons coming from the central star \citep{Dopita06c}. This  $\sim 20$\mum\ peak helps to distinguish UC \HII regions from the  the compactness parameter, since they change the slope in the $1.2 < \log (\lambda / \mum) <1.5$ wavelength region, but make little difference to the SED in the 10\mum ``PAH gap'' (\emph{see} Figure (3) panel (d)). We can determine $F_{\rm UCHII}$  to an accuracy of $\pm0.2$.

The presence of an active galactic nucleus (AGN) may sometimes be confused with a large fraction of UC \HII regions, since both contain very hot dust giving a strong $\sim 20$\mum\ peak. However, the presence of an AGN reveals itself in other ways, such as through a strong featureless continuum which washes out the PAH features, and through the high-excitation emission lines which appear in the IRS spectra. This issue will be further discussed below.

\subsection{Dust Attenuation}
Very high dust attenuation, equivalent to several magnitudes of optical extinction are required before observable effects become visible at IRAC and IRS wavelengths. Thus, the accuracy with which we can determine the foreground or the starburst reddening are dependent upon the absolute values of these quantities. This is illustrated in Figure (4) in which three separate cases are shown. We conclude that when $A({\rm V})_{\rm SB} < 3$mag. the attenuation is essentially unmeasurable. When $ A({\rm V})_{\rm SB} >  5$mag the attenuation can be estimated to an accuracy of $ \pm  5$mag. Note the flattening in the color index between he IRAC bands at 3.6\mum\ and 4.5\mum\  for the most heavily reddened values. This can be used to estimate the foreground reddening in those objects with $A({\rm v})_{\rm old} > 5$mag. However, the optical and UV SEDs are very much more sensitive to reddening, and these should be used in preference, if available.

\section{Results}

\begin{figure*}
\includegraphics[width=\hsize]{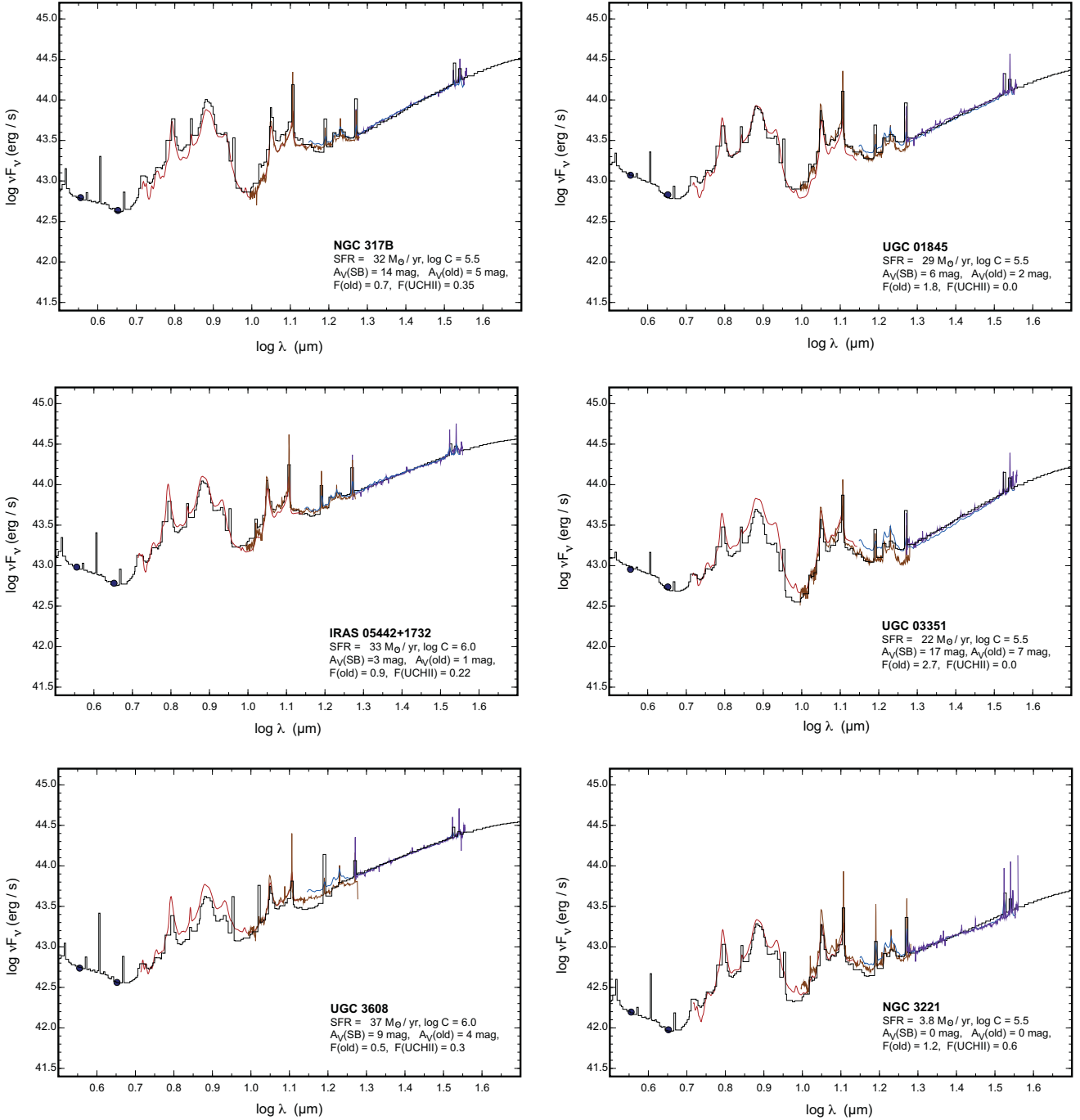}
\caption{The fit to the observed SEDs of GOALS starburst galaxies. The colored lines represent the IRS spectra, the dark blue points the IRAC photometry points, and the stepped black line is our SED model fit given for each of the wavelengths computed in our model. Note that the issues associated with the normalization of the spectra affect the goodness of fit for some objects (for example, UGC3608 and UGC03351 on this panel, and NGC5936, IRAS F18292-3513 and ESO 286-G035 on subsequent panels).}\label{fig_SED_fit}
\end{figure*}

The results of our fitting for the 16 objects which could be well-described by the ``pure starburst'' modeling are shown in table \ref{tab:Fits} and figures  \ref{fig_SED_fit} \emph{et seq.} Note that all of these are rather good fits to the observed spectra, including the emission lines, although we would need to have chemical abundances derived from optical spectra of these objects in order to better quantify this statement.

\begin{figure*}
\includegraphics[width=\hsize]{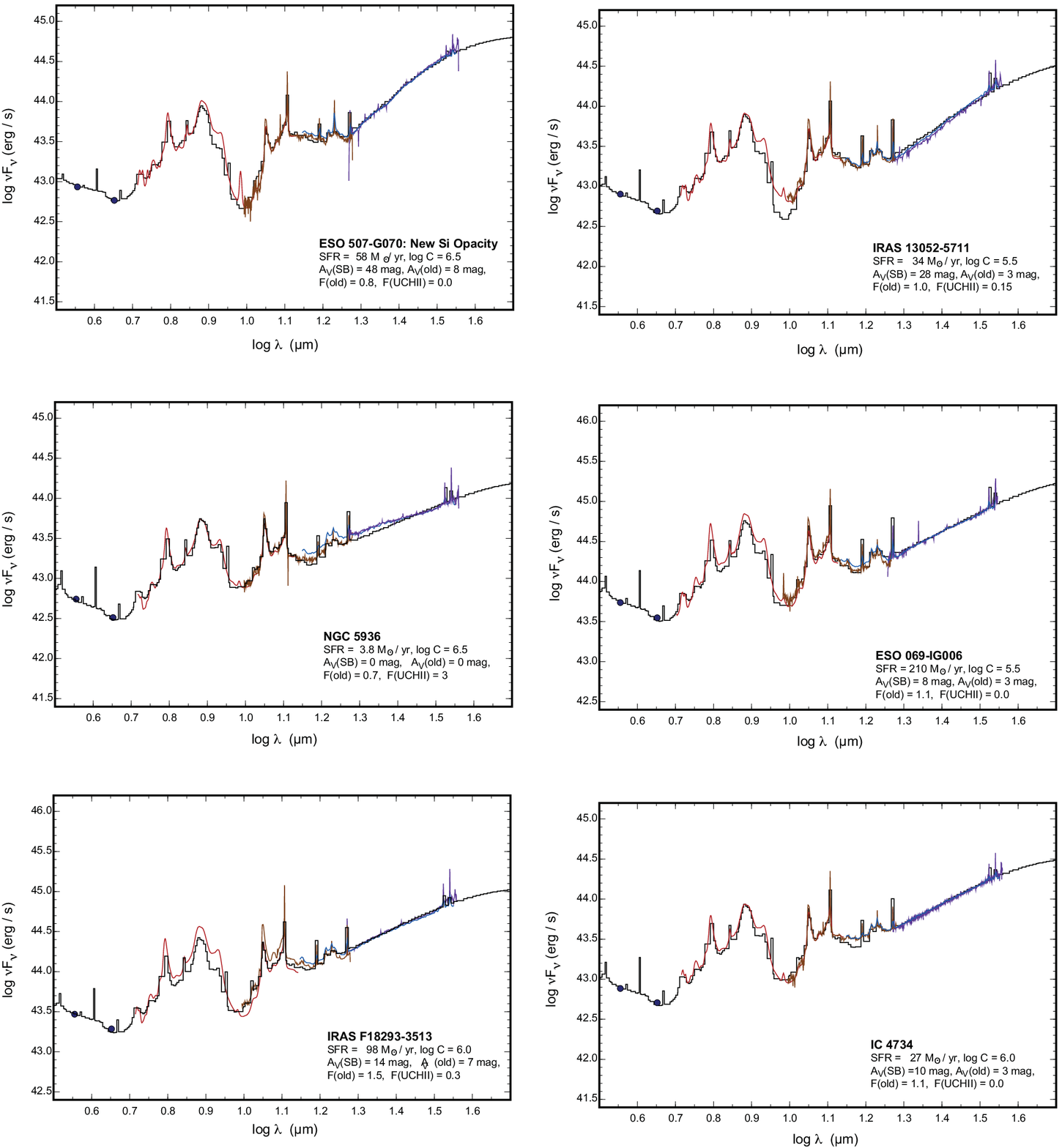}
\caption{As figure \ref{fig_SED_fit}.}
\end{figure*}
\begin{figure*}
\includegraphics[width=\hsize]{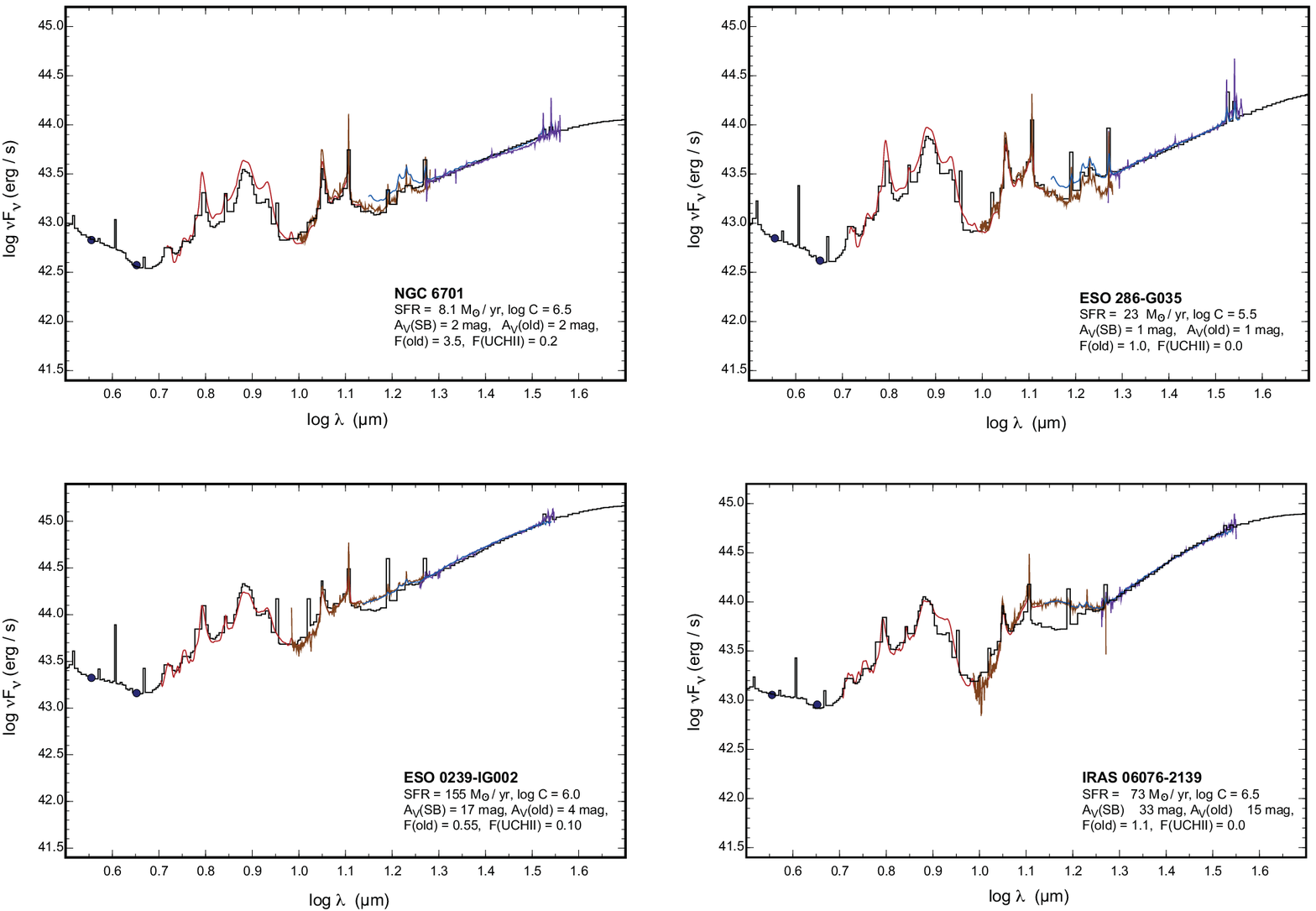}
\caption{As figure \ref{fig_SED_fit}.}
\end{figure*}

As might be expected for LIRGs, all of our galaxies turn out to have high compactness parameters ($ 5.5 < \log{\cal C} < 6.5$), indicating high pressures in the star formation region. From the definition of the compactness parameter given in Eqn. (1), if the mean cluster mass is $\log M_{\rm cl} \sim 3 \times10^5 M_{\odot}$, then $5.5 < \log(P/k) < 8.0$ (K~cm$^{-3}$). All of these objects are dense, high-pressure star formation regions with relatively ``hot'' dust. Interestingly, the inferred star formation rate is not correlated with the compactness of the starburst for this sample. 

We would not necessarily expect to find such a correlation, because compactness is description of the geometrical relationship between the dust and the stars which heat it, while star formation rate is a global property of the galaxy. However, statistically for starburst galaxies a correlation between these quantities should exist, and indeed, is hard-wired into the \citet{Dale02} template fitting. The physical reason for a correlation is the existence of \citet{Kennicutt98} star formation law. This shows that the surface rate of star formation $\Sigma_{\rm SFR}$ is related to the surface density of gas  $\Sigma_{\rm gas}$ through the relation  $\Sigma_{\rm SFR} \propto  \Sigma_{\rm gas}^{1.5}$. Thus object with high rates of star formation tend to be more compact, except in the cases where the star formation region is greatly extended. Indeed, \citet{Diaz-Santos09} has recently found that in many LIRGs the extended component of their MIR continuum emission originates in scales up to 10 kpc, and may contribute as much as the nuclear region to their total MIR luminosity. In addition, the more extended a galaxy is in the mid-IR the cooler far-IR colors it has. The absence of any correlation between star formation rate and compactness in our sample is probably more related to our selection for sources which are obviously compact, predicated by our need to start modeling objects which are morphologically simple.

Although the objects show a great variety in the attenuation associated with the starburst ($0 <  A({\rm v})_{\rm SB} < 48$ mag), the general extinction associated with the older population of stars (derived mainly from the color of the two IRAC photometry points, as described above) is much lower  ($0 <  A({\rm v})_{\rm gal} < 8$ mag). This is consistent with our expectation that regions of high specific star formation rate have high optical depths to dust absorption.

The old stars fraction lies in the range  $0.5 < F_{\rm old} < 3.5$, consistent with starburst ages up to a few $10^8$yr, similar to the dynamical timescales of galaxies. The highest value is found in NGC~6701, which has low reddening of both the starburst, and in the foreground. 

The fraction, $F_{\rm UCHII}$, of young ($< 1.0$Myr) ultra-compact \HII\  regions is relatively low in most objects ( $0.0 < F_{\rm UCHII}< 0.35$), consistent with the expected lifetimes of the ultra-compact phase of evolution ($<1-2$Myr).  $F_{\rm UCHII}$ reaches large values in only two objects, NGC~3221 ($F_{\rm UCHII} \sim 0.6$) and NGC~5936($F_{\rm UCHII} \sim 3$). It is possible that both of these objects may have a low-level AGN contribution, since AGNs also produce hot dust as described above, and it would therefore be difficult to distinguish from the contribution of UCHII regions in the $\sim 20$\mum\ region of the spectrum. NGC~3221 is nearly edge-on, so any AGN would be difficult to detect. However, a fairly bright and compact radio source was observed by \citet{Irwin00}, and an X-ray detection was made by \citet{Rephaeli95}, suggesting that there may be a buried AGN in this object. However, NGC~5936 is a nearly face-on galaxy with a circum-nuclear stellar ring, and shows no sign of X-ray emission \citep{Rephaeli95}, so this appears to be a bona-fide starburst with many single OB stars rather than having the star formation concentrated into clusters.  $F_{\rm UCHII} \sim 3$ implies that the total current star formation rate in this system is $4\times 3.8 \sim 15$\Msunpyr.

In table \ref{tab:Fits} we have given the star formation rates inferred by \citet{Howell10} for those galaxies which are in common. All our estimated star formation rates tend to be systematically lower, and for a few cases, the star formation rates we have derived are dramatically lower. The cause of this discrepancy is purely due to aperture effects. We have extracted spectra only from the IRS aperture, not the whole galaxy as is the case in the \citet{Howell10} study. We must also bear in mind that the total instantaneous star formation rates for some galaxies are higher when we include the contribution of the UCHII regions. For example, including the UCHII regions gives a total star formation rate for NGC~5936 of $15$\Msunpyr, as noted above, much closer to the $25.23$\Msunpyr inferred by \citet{Howell10}. Likewise, the total star formation rate in NGC~3221 is increased to $6.1$\Msunpyr by the addition of the UCHII component.

Finally, we want to draw attention to the somewhat peculiar object IRAS 06076-2139, shown in the last panel of Figure 8. Whilst the majority of the SED is well fit by our starburst model, there is a strong excess in the PAH emission at about 12.5\mum. This feature has been seen in spectral maps of NGC~7027, where it has been interpreted as due to (uncharged) PAH clusters \citep{Rapacioli05,Berne08,Tielens08}. This type of PAH spectrum has been observed in other ULIRGS, for example IRAS 13416-6243, which has been dubbed a ``class C'' source by \citet{Tielens08}. IRAS 06076-2139 was the subject of a detailed integral field study by \citet{Arribas08}. This system is a very close double system, with relative velocities $>500$km~s$^{-1}$, which have recently undergone a strong collision. This collision, and the resultant shocks may have something to do with the peculiar PAH spectrum.

\subsection{Objects with Embedded AGNs}
Of the 19 objects selected we found clear evidence for embedded AGN in only three objects. This is a surprisingly low fraction for such a luminous set of objects, but this simply reflects the pre-selection of sources towards those that contained strong PAH features which are likely to be modeled by simple starbursts. It will need a more complete and unbiased sample to determine a reliable starburst / AGN fraction. In our sample, the SED of MCG-03-34-064 is completely dominated by the bolometric luminosity of the AGN, and both the continuum associated with hot dust and the strong high-excitation emission lines expected in an AGN are clearly seen, \emph{see} Fig 8.

Two other objects show a mixed starburst + AGN excitation, IRAS 08355-4944 and IRAS 23436+5257 (\emph{see} Fig 9). Both of these objects show an excess in the IRAC 4.5\mum\ band, and this excess extends all the way up to at least 20\mum.  In these objects the PAH band strengths are suppressed, presumably by the underlying hot dust continuum. There is also clear evidence of the high excitation lines expected in objects containing and AGN. We have estimated a star formation rate of $\sim 85$ \Msunpyr\ in IRAS 08355-4944, and  $\sim 50$ \Msunpyr\ in  IRAS 23436+5257, but both of these are overestimates of the true star formation rate, as the contribution of the AGN to the bolometric luminosity cannot be properly disentangled in these objects. One possible way would to use only PAH fitting to estimate the underlying star formation, and then to add an AGN SED to the intrinsic starburst SED. This was not done here because we do not yet have good AGN templates. This approach is being currently developed by \citet{Petric10}.

\begin{figure}
\includegraphics[width=\hsize]{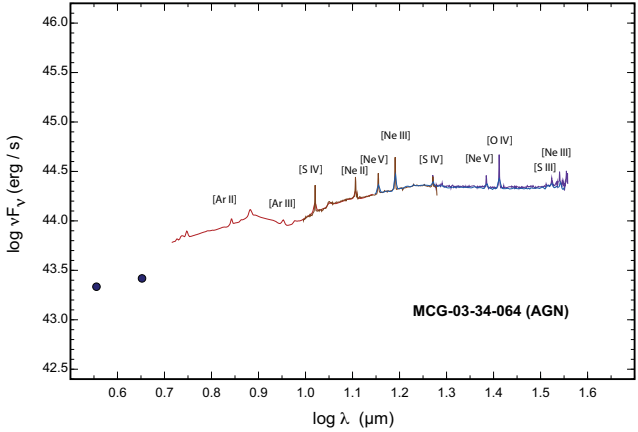}
\caption{MCG-03-34-064, an AGN-dominated SED. The principal line identifications are marked. note that the slope of the two IRAC photometry points is reversed, silicate absorption (if preset) is very weak, and that the PAH emission is practically swamped by the strong underlying continuum.} \label{fig_AGN}
\end{figure}

\begin{figure}
\includegraphics[width=\hsize]{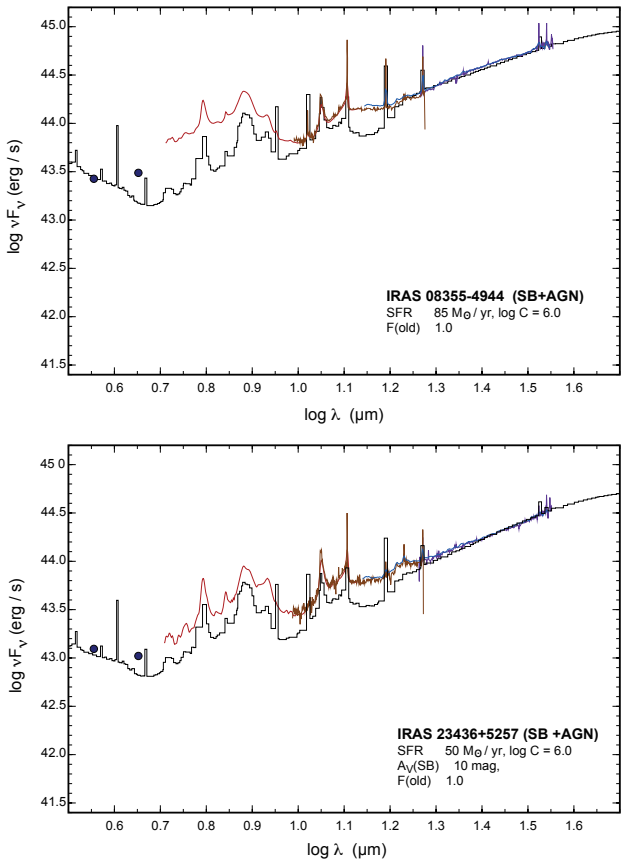}
\caption{The two examples of starburst + AGN SEDs found in our sample.} \label{fig_SB_AGN}
\end{figure}

\section{Conclusions}
In this paper we have shown that the IR SEDs of starburst galaxies in the GOALS sample observed with the \emph{Spitzer Space Telescope} can be modeled to a high degree of accuracy by the models which we have developed in a series of recent papers \citep{SED1,SED2,SED3,SED4}. The observed SEDs can be used with these models to constrain the absolute star formation rate, total extinction, and the $\sim 0-1, \sim 0-10$ and $\sim 10-100$Myr history of star formation. We can also determine a compactness parameter which is related to the pressure in the interstellar medium around the starburst. All of these sources have high compactness, and star formation rates are high, ranging up to 210 \Msunpyr\ .

We have also established that the silicate absorption in such dense starburst environments is fundamentally different than in the local ISM. Although this may be related to composition, we believe that it is more likely related to a population of grains with systematically larger sizes in the starburst environments.

\begin{acknowledgments}
Dopita acknowledges the continued support of the Australian Research Council (ARC) through Discovery  projects DP0984657 and DP0664434.  Dopita \& Armus thank the Aspen Center for Physics for providing the stimulus for this collaborative research. This research has made use of the NASA/IPAC Extragalactic Database (NED) which is operated by the Jet Propulsion Laboratory, California Institute of Technology, under contract with the National Aeronautics and Space Administration. We thank the anonymous referee for the great care and knowledge demonstrated in the referee reports, which has greatly improved our resultant paper.
\end{acknowledgments}


\clearpage

\begin{table*}[t]
\caption{Starburst Silicate Absorption compared with the Astronomical Silicate absorption in the 7.5-40 $\mu$m wavelength region. The column  D\&L84 is the attenuation of flux at wavelength $\lambda$ for $A_{\rm V}=1.0$mag using the \citet{DL84} `astronomical silicates'. The column labelled ``Starburst Si'' is the attenuation with our modified silicate absorption curve. The normalization of the two curves is identical at the peak of the 10\mum\ feature. \newline} \label{tab:Si-abs}
\small
\begin{tabular}{@{}cccccc@{}}
\tableline \tableline
{$\lambda$~($\mu$m)} & {D\&L84} & {Starburst Si} & {$\lambda$~($\mu$m)} & {D\&L84} & {Starburst Si}\\
\tableline
7.588   & 2.211E-02 &   2.211E-02 &   16.260  & 1.911E-02 &   2.957E-02 \\
7.682   & 2.280E-02 &   2.280E-02 &   16.590  & 1.958E-02 &   3.065E-02 \\
7.762   & 2.343E-02 &   2.343E-02 &   16.920  & 2.006E-02 &   3.177E-02 \\
7.856   & 2.412E-02 &   2.412E-02 &   17.260  & 2.048E-02 &   3.283E-02 \\
7.951   & 2.520E-02 &   2.469E-02 &   17.540  & 2.081E-02 &   3.375E-02 \\
8.023   & 2.836E-02 &   2.666E-02 &   17.780  & 2.098E-02 &   3.443E-02 \\
8.124   & 3.383E-02 &   3.011E-02 &   18.080  & 2.090E-02 &   3.470E-02 \\
8.333   & 3.739E-02 &   3.280E-02 &   18.130  & 2.081E-02 &   3.495E-02 \\
8.500   & 4.015E-02 &   3.530E-02 &   18.440  & 2.066E-02 &   3.512E-02 \\
8.637   & 4.373E-02 &   3.739E-02 &   18.530  & 2.042E-02 &   3.471E-02 \\
8.750   & 4.719E-02 &   4.015E-02 &   18.840  & 1.967E-02 &   3.385E-02 \\
8.947   & 4.893E-02 &   4.373E-02 &   19.000  & 1.881E-02 &   3.218E-02 \\
9.065   & 5.331E-02 &   4.719E-02 &   19.750  & 1.823E-02 &   3.099E-02 \\
9.118   & 5.656E-02 &   5.193E-02 &   20.150  & 1.764E-02 &   2.998E-02 \\
9.444   & 5.386E-02 &   5.631E-02 &   20.500  & 1.690E-02 &   2.872E-02 \\
9.571   & 5.175E-02 &   5.656E-02 &   21.210  & 1.616E-02 &   2.746E-02 \\
9.962   & 5.079E-02 &   5.482E-02 &   21.630  & 1.578E-02 &   2.683E-02 \\
10.000  & 4.918E-02 &   5.250E-02 &   22.020  & 1.548E-02 &   2.632E-02 \\
10.100  & 4.401E-02 &   4.818E-02 &   22.100  & 1.515E-02 &   2.575E-02 \\
10.450  & 4.280E-02 &   4.544E-02 &   22.500  & 1.485E-02 &   2.524E-02 \\
10.600  & 4.130E-02 &   4.436E-02 &   22.690  & 1.427E-02 &   2.426E-02 \\
10.630  & 4.003E-02 &   4.351E-02 &   23.110  & 1.375E-02 &   2.337E-02 \\
10.780  & 3.652E-02 &   4.064E-02 &   23.780  & 1.348E-02 &   2.301E-02 \\
10.780  & 3.885E-02 &   4.273E-02 &   24.120  & 1.321E-02 &   2.214E-02 \\
10.940  & 3.504E-02 &   3.946E-02 &   24.380  & 1.298E-02 &   2.133E-02 \\
11.140  & 3.433E-02 &   3.912E-02 &   24.740  & 1.271E-02 &   2.050E-02 \\
11.200  & 3.310E-02 &   3.817E-02 &   24.970  & 1.252E-02 &   1.981E-02 \\
11.280  & 3.123E-02 &   3.643E-02 &   25.440  & 1.237E-02 &   1.920E-02 \\
11.410  & 2.942E-02 &   3.473E-02 &   25.480  & 1.211E-02 &   1.845E-02 \\ 
11.560  & 2.789E-02 &   3.331E-02 &   25.850  & 1.171E-02 &   1.750E-02 \\
11.700  & 2.688E-02 &   3.248E-02 &   26.230  & 1.129E-02 &   1.655E-02 \\
11.870  & 2.573E-02 &   3.146E-02 &   27.090  & 1.099E-02 &   1.580E-02 \\
11.920  & 2.404E-02 &   2.975E-02 &   27.630  & 1.069E-02 &   1.508E-02 \\
12.100  & 2.279E-02 &   2.852E-02 &   28.180  & 1.041E-02 &   1.441E-02 \\
12.280  & 2.210E-02 &   2.800E-02 &   28.750  & 1.012E-02 &   1.374E-02 \\
12.400  & 2.098E-02 &   2.689E-02 &   29.310  & 9.772E-03 &   1.302E-02 \\
12.460  & 1.951E-02 &   2.530E-02 &   30.000  & 9.460E-03 &   1.236E-02 \\
12.720  & 1.864E-02 &   2.445E-02 &   30.850  & 9.209E-03 &   1.181E-02 \\
12.910  & 1.780E-02 &   2.364E-02 &   31.470  & 9.040E-03 &   1.137E-02 \\
12.990  & 1.671E-02 &   2.245E-02 &   32.100  & 8.879E-03 &   1.096E-02 \\
13.190  & 1.592E-02 &   2.164E-02 &   32.320  & 8.673E-03 &   1.050E-02 \\
13.390  & 1.486E-02 &   2.044E-02 &   32.960  & 8.505E-03 &   1.010E-02 \\
13.600  & 1.416E-02 &   1.971E-02 &   33.450  & 8.324E-03 &   9.700E-03 \\
14.090  & 1.436E-02 &   2.022E-02 &   33.920  & 8.134E-03 &   9.298E-03 \\
14.300  & 1.458E-02 &   2.077E-02 &   34.500  & 7.948E-03 &   8.914E-03 \\
14.510  & 1.481E-02 &   2.135E-02 &   35.070  & 7.740E-03 &   8.516E-03 \\
14.690  & 1.499E-02 &   2.187E-02 &   35.680  & 7.414E-03 &   8.004E-03  \\
14.920  & 1.552E-02 &   2.291E-02 &   36.390  & 7.083E-03 &   7.501E-03 \\
15.000  & 1.625E-02 &   2.400E-02 &   37.860  & 6.782E-03 &   7.047E-03 \\
15.430  & 1.735E-02 &   2.622E-02 &   38.660  & 6.490E-03 &   6.615E-03  \\
15.680  & 1.849E-02 &   2.780E-02 &   40.000  & 6.249E-03 &   6.249E-03  \\
\tableline
\end{tabular}
\end{table*}
\clearpage

\begin{table*}[t]
\caption{Fitted parameters for the `pure' starburst galaxies. {SFR}$_{int}$ is the integrated star formation rate as derived for the galaxies in common with our sample by \citet{Howell10} \newline} \label{tab:Fits}
\small
\begin{tabular}{@{}llccccccc@{}}
\tableline \tableline
{Position:} & {Name} & {SFR}$_{int}$ &{SFR} & {$\log C$} & {$A_{\rm V}$(SB)} & {$A_{\rm V}$(old)} & {F(old)} & {F(UCHII)} \\
{RA (J2000.0) Dec}&{}&{(\Msunpyr)}&{(\Msunpyr)}& &{(mag.)}&{(mag.)}&{}&{} \\
\tableline
00:57:40 +43:47:32 & NGC 317B & 26.97 & 32 & 5.5 & 14 & 5 & 0.7 & 0.35 \\
02:24:08 +47:48:11 & UGC 01845 & & 29 & 5.5 & 6 & 2 & 1.8 & 0.0 \\
05:47:11 +17:34:37  & IRAS 05442+1732 & & 33 & 6.0 & 3 & 1 & 0.9 & 0.22 \\
05:45:48 +58:42:04  & UGC 03351 &  & 22 & 5.5 & 17 & 7 & 2.7 & 0.0 \\
06 :09:46 -21:40:24 & IRAS 06076-2139 & & 73 & 6.5 & 33 & 15 &1.1 & 0.0 \\
06:57:34 +46:24:11 & UGC 03608 & & 37 & 6.0 & 9 & 4 & 0.5 & 0.3 \\
10:22:20 +21:34:10 & NGC 3221 & 21.75 & 3.8 & 5.5 & 0 & 0 & 1.2 & 0.6 \\
13:02:52 -23:55:18 & ESO 507-G070 & 62.77& 58 & 6.5 & 48 & 8 & 0.8 & 0.0 \\
13:08:18 -57:27:30 & IRAS 13052-5711 & & 34 & 5.5 & 28 & 3 & 1.0 & 0.15 \\
15:30:01 +12:59:22 & NGC 5936 & 25.23 & 3.8 & 6.5 & 0 & 0 & 0.7 & 3.0 \\
16:38:13 -68:26:43 & ESO 069-IG006 & & 210 & 5.5 & 8 & 3 & 1.1 & 0.0 \\
18:32:41 -34:11:27  & IRAS F18293-3514 & & 98 & 6.0 & 14 & 7 & 1.5 & 0.3 \\
18:38:26 -57:29:26 & IC 4734 & 39.30 & 27 & 6.0 & 10 & 3 & 1.1 & 0.0 \\
18:43:12 +60:39:12 & NGC 6701 & & 8.1 & 6.5 & 2 & 2 & 3.5 & 0.2 \\
21:04:11 -43:35:33 & ESO 286-G035 & 27.58 & 23 & 5.5 & 1 & 1 & 1 & 0.0 \\
22:49:40 -48:50:58 & ESO 239-IG002 & & 155 & 6.0 & 17 & 4 & 0.5 & 0.1 \\
\\
\tableline
\end{tabular}
\end{table*}
\clearpage

\end{document}